\documentclass[useAMS]{mn2e}
\input epsf
\usepackage{lscape}
\usepackage{rotating}
\usepackage{times}

\def\two{\,{\sc ii}}

\input psfig.sty
\begin{document}

\title[Mid Infra Red Imaging]{{\bf MSX} Mid infra red imaging of massive
star birth  environments. II: Giant H\,{\sc ii} regions}

\author[Peter S. Conti and Paul A. Crowther] {Peter S. Conti
\thanks{E-mail: pconti@jila.colorado.edu (PSC);
Paul.Crowther@sheffield.ac.uk (PAC)},  and  Paul A.
Crowther\\ 
JILA and APS Department, University of Colorado, Boulder CO 80309-0440, 
USA\\Department of Physics and
Astronomy, University of Sheffield, Hicks Building, Hounsfield Rd,
Shefffield, S3 7RH, UK}
\date{Accepted. Received; in original form}

\pagerange{\pageref{firstpage}--\pageref{lastpage}}
\pubyear{2004}

\maketitle

\label{firstpage}

\begin{abstract} 
We conduct a Galactic census of Giant H\,{\sc ii} regions,
based on the all sky 6cm dataset  of Kuchar \& Clark, plus the
kinematic distances obtained by Russeil. From an inspection of 
mid-IR Mid-course Space Experiment ($MSX$) and far-IR $IRAS$ Sky Survey Atlas
images we identify a total of 56 GH\,{\sc ii} regions in the Milky Way,
of which 15\% (65\%) can be seen at optical (near-IR) wavelengths.
The mid to far-IR fluxes from each GH\,{\sc ii} region are measured, and
sample the  thermal emission from the ubiquitous dust
present within the exciting clusters of OB stars, 
arising from the integrated luminosity of the hot stars heating the 
cluster dust, for which we obtain $\log L$(IR)=5.5--7.3$L_{\odot}$.
The mid-IR 21$\mu$m 
spatial morphology is presented for each GH\,{\sc ii} region, and
often indicates multiple emission sources, 
suggesting complicated cluster formation.  
IR colour-colour diagrams are presented, providing information about the
temperature distribution and optical depth of the dust. For the
clusters of our study, the dust is not optically thick to all the
stellar radiation, thus the measured infra-red luminosity is lower than
the L$_{\rm bol}$. As the dust environment of a  cluster begins to dissipate,
the thermal emission and its optical depth ought to decrease even before
the stars appreciably evolve.  We see evidence of this in our empirical
relationship between the integrated IR and Lyman continuum luminosities.
\end{abstract} 

\section{INTRODUCTION}

Massive stars, with O or early B spectral types on the main sequence are,
for the most part, born in giant molecular clouds (GMCs). Initial
``clumps'' made up of dense molecular gas along with appreciable
concentrations of dust comprise the birth material.  Once a luminous star
becomes sufficiently hot, substantial Lyman continuum luminosity is
produced and the surrounding hydrogen is rapidly ionized into a
Str\"{o}mgren Sphere. During these birth phases, the medium surrounding 
the stellar embryo is sufficiently dense ($\approx 10^6$cm$^{-3}$) that
the H\,{\sc ii} region radius is a small fraction of a parsec,
$\approx~$0.01--0.1 pc.  This ultra-compact H\,{\sc ii} (UCH\,{\sc ii})
region (e.g. Wood \& Churchwell 1989) is detectable at cm wavelengths. It
is surrounded by a natal dust ``cocoon'', which is somewhat opaque to 
(roughly) 
near infrared (NIR) and shorter wavelength radiation and so is heated by
the star.  These cocoons are observable as thermal sources in the mid 
infrared (MIR) and at longer wavelengths as has been shown previously
(e.g.  Crowther \& Conti 2003 - hereafter Paper~I). Further evolution of
the individual stellar birth cocoons proceeds as follows: The intense
radiation of the star (likely aided by its wind) dissipates  and
evacuates the dust that gradually expands. As it does so, its optical
depth diminishes and the OB type exciting star becomes revealed. The
UC\,{\sc ii} region also becomes larger, forming a compact H\,{\sc ii} and
finally a normal H\,{\sc ii} region.

For OB stars formed in a cluster, additional physics comes into play as 
the environment also contains gas and dust. Each luminous star affects
this cluster material by heating the dust and gradually dissipating it by
the strong stellar radiation fields and winds.  In addition, the Lyman
continuum luminosity from a large ensemble of OB stars can ionize a
substantial volume of space, resulting in an observable giant H\,{\sc ii}
(GH\,{\sc ii}) region. In a few cases (e.g. W49A) individual UCH\,{\sc 
ii} regions can contribute to the overall ionization of the cluster 
environment (e.g. Conti \& Blum 2002, Alves \& Homeier 2003).  
We would expect that a GH\,{\sc
ii} region would be ionized by OB stars which are, for the most part,
already beyond the UCH\,{\sc ii} stage (e.g. Blum et al. 2002). We are
interested in measuring parameters of the dust content of the OB star
clusters since that lifetime is much longer than that of the individual
natal dust cocoons surrounding the OB star members, but may still be
shorter than the stellar evolution time scale.

We began this study by considering those Galactic  (GH\,{\sc ii}) objects
that emit a Lyman continuum luminosity -- N(LyC) -- of at least $10^{50}$
photon\,s$^{-1}$, thus excited by more than ten O7V* stars (O star
equivalents - Vacca 1994).  Such regions ought to represent the main
locations of massive stars in our Galaxy. Following the technique
of Smith et al. (1978) we re-evaluate the census
of Galactic GH\,{\sc ii} regions  based on modern
radio surveys (e.g. Condon et al.1989) and  kinematical distances
(Russeil 2003), 
that is, they are derived from a Galactic rotation model and
H\,{\sc i} line velocity 
measurements.\footnote{The rotation model assumes axi-symmetric circular motions (no 
peculiar velocities).Inside the solar circle a near/far side ambiguity is 
present which must be resolved with other data.}
This GH\,{\sc ii}
sample has formed the basis of an ongoing systematic NIR imaging and 
spectroscopic study of the stellar content of these objects (e.g. Blum
2003 and \S~4).  It is already clear from our published and  unpublished
NIR imaging that many of the radio selected GH\,{\sc ii} sample contain
identified OB star clusters.

A primary objective of this paper is to study the newly selected
GH\,{\sc ii} regions in
the MIR with $MSX$ at a higher (circular) spatial resolution than was possible with
$IRAS$
\footnote{The $IRAS$ central wavelengths of 12$\mu$m, 25$\mu$m, 60$\mu$m
and 100$\mu$m have rectangular spatial resolutions of 30$''$, 30$''$,
1$'$ and 2$'$, respectively.}. Kraemer et al. (2003) present studies of
nearby star forming regions (Orion, Pleiades etc.) with $MSX$, whilst
{\it Spitzer} has already began producing yet higher spatial resolution MIR
maps (Churchwell et al. 2004; Whitney et al. 2004). Nevertheless,
the present work  extends previous studies to a complete census of
the most luminous radio selected, massive star forming regions. In  many cases 
the spatial morphology is quite complicated, indicating the presence of more
than one dust emission source, so previous H\,{\sc ii} region studies
making use of the $IRAS$ Point Source Catalogue (e.g. Codella et al.
1994) may be inappropriate due to the widely variable spatial scale of
Galactic GH\,{\sc ii} regions (Chan \& Fich 1995). We will make
measurements of the various IR fluxes through the $IRAS$ filters
using the Sky Survey Atlas (ISSA).

We would like to compare measurements of the thermal emission from the
integrated cluster dust with the radio free-free emission from the  hot
stars within. The former depends upon the far UV (FUV) luminosity of the
ensemble of OB stars in the cluster(s) and the extent and temperature
distribution of the dust within it.  The latter depends upon the 
integrated extreme UV (EUV) luminosity of these same stars. In our
Paper~I we have proposed that these same measurements fit reasonably
well a simple spherical model for the (mostly single star excited)
UCH\,{\sc ii}  regions. We will revisit this relationship below.

It is well known that UCH\,{\sc ii} regions have spectral energy
distributions (SEDs) which are not fit by a single temperature black body
(e.g. Wolfire \& Churchwell 1994). Rather the observed SEDs of UCH\,{\sc
ii} regions are broad with a peak near 100$\mu$m, corresponding to a dust
temperature of 30 deg K. However, other hotter material is also present.  
What is the situation for the dust in those OB star clusters with dust?  
Here MIR and far-IR (FIR) photometry will help constrain the SEDs of dusty
OB clusters.
  
In Section~\ref{sam} we present our revised GH\,{\sc ii} catalogue,
which is complemented by the largest GH\,{\sc ii} regions in the 
Magellanic 
Clouds. The procedure for
analysing the observations is contained in Section~\ref{obs}. We discuss 
in Section~\ref{morph} the individual MIR images of this set of data
which are presented in the Appendix. In Section~\ref{flux} we discuss the
similarities and differences among the objects using different IR flux
ratios (colours). We compare the GH\,{\sc ii}/H\,{\sc ii} colours with
those of the UCH\,{\sc ii} regions from Paper I. We also consider the
relationships between the IR luminosity and the radio fluxes for all
the data. A discussion is given in Section~\ref{summ}.

\begin{figure} 
\epsfxsize=8.8cm\epsfbox[0 50 504 740]{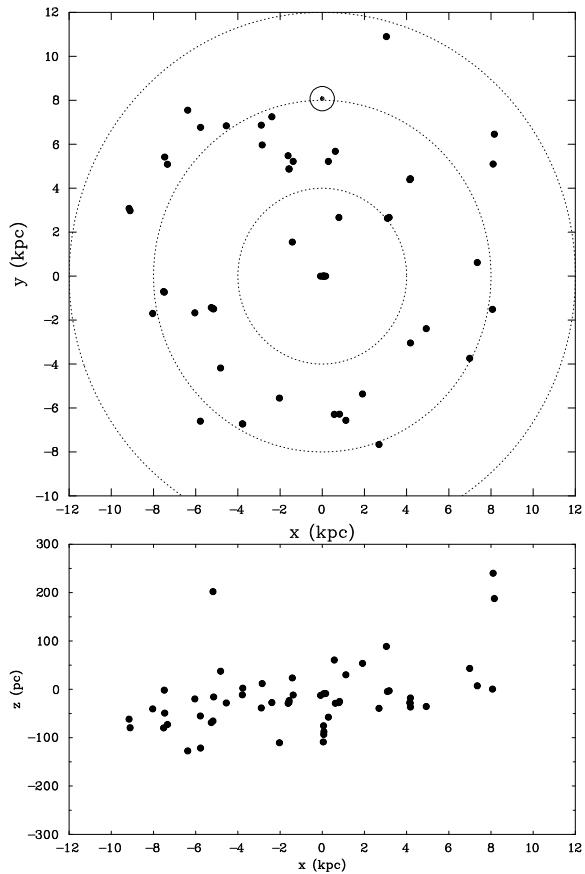}     
\caption{Distribution of GH\,{\sc ii} regions (filled circles) 
in the Galactic disk  used in this study. 
The location of the Sun is indicated at a Galactocentric distance
of 8kpc (Reid 1993), as are 
radii of 4, 8  and 12 kpc (dotted lines).}
\label{galaxy} 
\end{figure} 

\section{SELECTION OF SAMPLE}\label{sam}

In this paper we look at a planopy of massive star birth environments 
in multiple excited GH\,{\sc ii} regions. At an initial stage of this 
study we
based our sample on forty GH\,{\sc ii} regions of Blum (2003). This sample
was directly drawn from Smith et al. (1978),
which was  based on low spatial resolution 6cm radio fluxes
taken from Reifenstein et al. (1970) and Wilson et al. (1970), except that
a revised Solar galactocentric distance of 8kpc was adopted (Reid 1993).

In preference
to these older compilations, we used the higher spatial resolution 6cm
study of H\,{\sc ii} regions by Kuchar \& Clark (1997), based on the NRAO 91m 
Condon et al. (1989) survey plus the  ATNF 64m (Parkes) Condon et al. (1993)
 survey. These were used in preference to (admittedly yet higher) 
spatial resolution radio data such as the NRAO VLA Sky Survey (NVSS) at 1.4GHz, 
since we required all sky coverage 
at an optically thin radio frequency, for which 5GHz is superior to 1.4GHz.
In two cases (M17 and W43) we used 6cm radio fluxes from Downes et al. (1980),
since these were missing from the Kuchar \& Clark (1997) compilation, presumably
due to saturation effects.

From our initial list of candidate GH\,{\sc ii} regions based solely upon
Russeil (2003) radio derived kinematical distances, we first revised distances 
for several well-known regions based on more recent radio studies by Araya et
al. (2002), Watson et al. (2003), Sewilo et al. (2004) and Corbel \&
Eikenberry (2004). This revision led several candidates to be revised to
the near (rather than far)  distance, demoting them to normal H\,{\sc ii}
regions, including G34.255+0.144 (Araya et al. 2002), G10.617--0.384 and
G10.664--0.467 (Corbel \& Eikenberry 2004). 

Subsequently, further candidates were rejected on the basis of spatial location
in the Milky Way inferred from the kinematical distance assumed.  Two were rejected
since their Galactocentric distance was predicted to exceed 12kpc (based on a
Solar Galactocentric distance of 8kpc, Reid 1993), namely G7.472+0.060 and
G173.599+2.798. In addition, two further candidates were omitted since
their `far' kinematical distance implied a distance of z$\sim$300\,pc from
the Galactic plane, namely G351.613--1.270 and G351.695--1.165. It is
likely that these two sources lie on the 'near' side of the Galactic
Centre, such that they are normal H\,{\sc ii} regions. Two other
candidates lying $z\sim$200 pc from the Galactic plane {\it are} included
(G70.300--1.600, G79.293+1.296) since they lie outside the Solar circle,
and their distances are probably genuine. 

Finally, ten further candidates,
some associated with the W49 and W51 complexes, G2.901--0.006, 
G13.381+0.071,
G26.103--0.069, G289.760-1.155, G338.086+0.007, G342.382--0.044 were also
dropped since their 21$\mu$m fluxes were less than 5\% of that for the
mean GH\,{\sc ii} region, after being adjusted to a uniform distance in
all cases. Finally, we also omitted the Galactic Centre cluster
(G359.951--0.037) since we considered a study of its IR/radio
properties might be affected by unusual conditions relative to normal
star forming regions.

Table 1 gives the listing of the remaining 56 GH\,{\sc ii} regions as
derived from radio measurements. Figure~\ref{galaxy} indicates the
Galactic distribution of GH\,{\sc ii} regions in Galaxy, as viewed from
above, projected onto the disk, and along the disk of the Milky Way.  
This is analogous to Fig.~11 from Georgelin \& Georgelin (1976), albeit
based solely on radio selected GH\,{\sc ii} regions. 
From the original Smith et al. (1978) sample, there was a lack of radio GH\,{\sc ii}
regions on the far side of the Galaxy. It is apparent that more
contemporary distance estimates has removed this puzzle.  The spiral arm
structure is not immediately apparent (see Russeil 2003), although any
interpretations must be treated with caution as the distances have been
determined kinematically. The handful of objects so far studied
photometrically and spectroscopically in the NIR (e.g. see Blum et al.
2002) have spectroscopic parallaxes which move them closer to the Sun, 
two substantially. 
The appearance of this figure might change dramatically once
more of the OB star clusters exciting the GH\,{\sc ii} have been
identified and their distances determined by this method. With the
exception of several sources at the Galactic Centre itself, there is only
one region apparently within a galactocentric distance of 4kpc, namely
G347.611+0.204.

\begin{small}
\begin{table*}
\caption{Catalogue of Galactic GH\,{\sc ii} regions based on 6cm fluxes 
of Kuchar \& Clark (1997) and distances from Russeil (2003), except where
noted. Lyman continuum fluxes are corrected for dust extinction, following 
Smith et al. (1978) from which we also derive  $N_{e}$. We indicate whether
the GH\,{\sc ii} region is seen
optically ($\bullet$) or in the NIR ($\circ$).}
\begin{tabular}
{l@{\hspace{2mm}}r@{\hspace{1.5mm}}r@{\hspace{2.5mm}}
r@{\hspace{2.5mm}}r@{\hspace{2.5mm}}
r@{\hspace{2.5mm}}r@{\hspace{2.5mm}}r
@{\hspace{2.5mm}}r@{\hspace{2.5mm}}r@{\hspace{2mm}}
r@{\hspace{2.5mm}}r@{\hspace{2.5mm}}l@{\hspace{3mm}}l
}
\hline
Name & $l$ & $b$ & $d$ & Ref & F(Jy) & Ref & $\theta$($'$) & $N_{e}$   & $T_{e}$ & Ref & 
N(LyC)&Opt/\\
     &     &     & kpc &     &  6cm  &     & 6cm           & cm$^{-3}$ & 10$^{3}$K &   & 
log(s$^{-1}$)&NIR?\\
\hline
   & 0.361 & -0.780 & 8.0 & 7 & 7.4 & 5 & 5.2 & 38 & 5.1 & 2& 50.02& \\
   & 0.394 & -0.540 & 8.0 & 7 & 8.8 & 5 & 5.6 & 40 & 7.5 & 2& 50.03& \\
   & 0.489 & -0.668 & 8.0 & 7 & 8.4 & 5 & 3.7 & 69 & 5.5 & 2& 50.12& \\
W24& 0.518 & -0.065 & 8.0 & 7 & 35.1& 5 & 3.7 & 148& 7.2 & 2& 50.91& \\ %*
   & 0.572 & -0.628 & 8.0 & 7 & 7.2 & 5 & 2.5 & 117& 6.2 & 2& 50.05& \\
Sgr~D
   & 1.149 & -0.062 & 8.0 & 7,11 &19.3 & 5 & 5.5 & 59 & 5.9 &14 & 50.52& $\circ$\\ %*
   & 2.303 & 0.243  &14.3 & 7    & 7.3 & 5 & 5.9 & 22 &  3.7 & 2& 50.45& \\
AMWW\,34& 3.270 & -0.101 & 14.3 & 7 & 9.9 & 5 & 5.7 & 31 & 7.2 &2 & 50.47& \\
AMWW\,35& 4.412 & 0.118 & 14.6 & 7 & 10.4 & 5 & 4.9 & 38 & 5.7 &2 & 50.56& \\
M8 &  5.973 & -1.178 &  2.8 & 7 & 113.4 & 5 & 7.5 & 158 & 7.7 &2 & 50.19& $\bullet$\\
   &  8.137 &  0.228 & 13.5 & 7 &   8.2 & 5 & 1.8 & 160 & 6.5 &2 & 50.47& $\circ$?\\
W31& 10.159 & -0.349 & 4.5 & 10 & 66.3 & 5 & 2.9 & 376 & 5.7 & 2& 50.66&$\circ$ \\
W31& 10.315 & -0.150 & 15 & 9,10 & 20.5 & 5 & 3.1 & 103 & 5.4 &2 & 50.90& $\circ$\\
M17& 15.032 & -0.687 & 2.4 & 7& 844.5 & 2 & 4.5 & 1033 & 9.1 & 2& 51.22&$\bullet$ \\
   & 20.733 & -0.087 & 11.8 & 7 & 19.5 & 5& 6.1 & 41 & 5.9 & 2& 50.72 &&\\
W42& 25.382 & -0.177 & 11.5 & 7 & 29.5 & 5 & 3.7 & 113 & 7.1 & 2& 50.93 &$\circ$\\
   & 29.944 & -0.042 &  6.2 & 7 & 25.5 & 5 & 3.7 & 140 & 6.1 & 2& 50.33& $\circ$\\
W43& 30.776 & -0.029 & 6.2 & 7 & 62.2 & 2 & 4.1 & 186 & 6.0 & 2& 50.83&$\circ$ \\
AMWW\,52 
    & 32.797 & 0.192 & 12.9 & 6 & 5.8 & 5& 2.3 & 102 & 9.5 & 2& 50.03& \\
W49A& 43.169 & 0.002 & 11.8 & 8 & 69 & 5& 3.0 & 243 & 8.7 &2 & 51.21&$\circ$ \\
   & 48.596 & 0.042 & 9.8 & 6 & 12.2 & 5& 4.2 & 65 & 7.8 & 2& 50.14& \\
W51& 48.930 & -0.286 & 5.5 & 7& 24.3 & 5& 4.4 & 117 & 8.2 & 2& 50.03& $\circ$\\
W51 West
   & 49.384 & -0.298 & 5.5 & 7 & 27.2 & 5 & 2.8 & 241 & 7.4 & 2& 50.15& $\circ$\\
W51A& 49.486 & -0.381 & 5.5 & 7 & 110.4 & 5& 2.8 & 446 & 7.0  & 2&  50.94&$\circ$ \\
W58A& 70.300 & 1.600 & 8.6 & 7& 13 & 5& 2.1 & 212 & 9.1 & 14& 50.06& $\circ$\\ %*
DR7& 79.293 & 1.296 & 8.3 & 7 & 15.8 & 5& 3.4 & 110 &6.9 & 14 & 50.15& $\circ$\\ 
   W3& 133.720 & 1.210 & 4.2 & 7& 74.7 & 5& 3.4 & 351 & 8.7 &1 & 50.25& $\circ$\\
RCW42& 274.013 & -1.141 & 6.4 & 7& 39.9 & 5& 2.9 & 259 & 7.9 & 4& 50.36&$\bullet$ \\
   & 282.023 & -1.180 & 5.9 & 7& 40.9 & 5& 3.8 & 175 & 6.2 & 4 &50.32& $\circ$\\
RCW49& 284.301 & -0.344 & 4.7 & 7& 263.2& 5 & 7.4 & 191 & 8.0 & 4& 50.96&$\bullet$ \\
NGC\,3372& 287.379 & -0.629 & 2.5 & 7& 145.6 & 5& 7.0 & 208 & 7.2 & 4& 50.11&$\bullet$ \\
& 289.066 & -0.357 & 7.9 & 7& 16.4 & 5& 6.0 & 51 & 8.5 & 4 &50.05& $\circ$\\
NGC\,3576& 291.284 & -0.712 & 3.1& 7 & 113 & 5& 2.5 & 776 & 7.5& 4& 50.28&$\bullet$ \\
NGC\,3603& 291.610 & -0.528 & 7.9& 7 & 261 & 5& 6.9 & 159 & 6.9& 4 & 51.50&$\bullet$ \\
& 298.227 & -0.340 & 10.4 & 7& 47.4 & 5& 3.8 & 150 & 8.6 & 4& 50.87& $\circ$\\
& 298.862 & -0.438 & 10.4 & 7 &42.4 & 5 &3.8 & 135 & 6.6 & 4&50.87& $\circ$\\
& 305.359 & 0.194  & 3.5  & 7& 56.4 & 5& 3.5 & 291 & 5.1 & 4& 50.13 & $\circ$\\
& 319.158 & -0.398 & 11.5 & 7&11.2 & 5& 5.9 & 34 & 6.3 &4& 50.30& \\
& 319.392 & -0.009 & 11.5 & 7& 8.9 & 5& 3.5 & 69 & 7.7 &4& 50.17 &$\circ$?\\
& 320.327 & -0.184 & 12.6 & 7 &6.3 & 5 &5.0 & 30 & 5.7 &4& 50.11 &$\circ$\\
RCW97& 327.304 & -0.552 & 3.0&7 & 64.9 & 5& 2.9 & 441 & 4.7 &4& 50.14 &$\bullet$\\
& 327.993 & -0.100 & 11.4 & 7& 5.9 &5& 2.7 & 79 & 6.0 &4& 50.08& $\circ$?\\
& 330.868 & -0.365 & 10.8 & 7&14.7 &5& 4.0 & 69 & 4.9 &4& 50.56& $\circ$?\\
& 331.324 & -0.348 & 10.8 & 7&6.5 & 5&1.8 & 143 & 3.5 &4& 50.28& $\circ$\\ %*
& 331.354 & 1.072 & 10.8 & 7&6.7 &5& 4.9 & 35 & 5.4 &4& 50.10& $\circ$\\
& 331.529 & -0.084 & 10.8 & 7&47.1 &5& 4.0 & 128 & 6.2& 4& 51.16 &$\circ$\\ %*
& 333.122 & -0.446 & 3.5 &7& 49.5 &5& 4.5 & 191 & 5.8 &4& 50.08& $\circ$\\
& 333.293 & -0.382 & 3.5 &7& 45.5 &5& 3.8 & 240 & 6.3 &4& 50.04& $\circ$\\
& 333.610 & -0.217 & 3.1 &7& 116.2 &5& 3.7 & 422 & 6.2& 4& 50.43&$\circ$ \\
& 338.398 & 0.164 & 13.1 &7& 25.7 &5& 5.5 & 54 & 6.6 &4& 50.90& \\
& 338.400 & -0.201 & 15.7 &7& 6.3 &5& 3.3 & 55 & 9.1 &4& 50.24& \\ %*
& 345.555 & -0.043 & 15.2 &7& 15.1 &5& 5.5 & 38 & 6.5 &4& 50.69& \\
& 345.645 & 0.009 & 15.2 &7& 11.2 &5& 4.2 & 51 & 7.8 &4& 50.53& \\
& 347.611 & 0.204 & 6.6 &7& 23.2 &5& 6.1 & 57 & 4.0 &4& 50.44& $\circ$\\
& 351.467 & -0.462 & 13.7 &7& 4.7 &5& 2.7 & 64 & 5.7 &4& 50.17& $\circ$\\
Sgr~C& 359.429 & -0.090 & 8.0&7 & 19.3 &5& 4.3 & 93 & 9.3 &4& 50.46& \\ %*
\hline
30 Dor& 279.448 & -31.678 & 50.0 & 12& 47.3 &5& 5.9 & 36 & 10.0&- & 52.32 &$\bullet$\\
NGC\,346& 302.100&-44.930 & 60.0 & 13 &2.7&3 & 3.0 & 22 & 10.0 &-& 51.05& $\bullet$\\
\hline
\end{tabular}  
\\
(1) Wilson et al. (1979); 
(2) Downes et al. (1980); (3) Israel (1980); (4) Caswell \& Haynes (1987); (5) 
Kuchar 
\& Clark (1997); 
(6) Araya et al. (2002); (7) Russeil (2003); (8) Watson et al. (2003); (9) Sewilo et al. 
(2004); (10) Corbel \& Eikenberry (2004); (11)
though see Blum \& Damineli (1999); (12) Madore \& Freedman (1998); 
(13) Harries et al.  (2003); (14) Wink et al. (1983)
\end{table*}
\end{small}

\section{MID AND FAR-IR OBSERVATIONS}\label{obs}

\subsection{$MSX$ Observations}

The Spatial Infrared Imaging Telescope (SPIRIT III) aboard 
the U.S. Department of Defense $MSX$ satellite surveyed the entire 
Galactic plane in four MIR spectral bands, named A, C, D, and E,
between 6 and 25$\mu$m at a ($\sim$3 pixel)  spatial resolution of
$\sim$18.3$''$.  A description of SPIRIT III
is given by Price et al. (2001). The entire area within
$\pm$4.5$^{\circ}$ of the Galactic plane was surveyed at least twice,
with four-fold coverage obtained in the first and fourth quadrants, and
to $\pm$3$^{\circ}$ for the remaining quadrants. The redundancy was
sufficient to permit combining the datasets onto a uniformly spaced grid,
such that the inherent spatial resolution of the instrument was
preserved. A total of 1680 95.1$'\times95.1'$ images were created in each
of the   spectral bands, spaced 90$'$ apart in each 
coordinate, providing 10$'$ overlap on adjacent images.  Each image
provides radiances on a grid in Galactic latitude and longitude, with a
grid spacing of 6$''$. The image units are of in-band radiance (W
m$^{-2}$ sr$^{-1}$).

Of the available bands, Band~A is the most sensitive, but its filter
footprint of 6.8--10.8$\mu$m includes the astrophysically strong, broad
silicate band at 9.7$\mu$m and the 7.7, 8.6$\mu$m PAH features (e.g. Faison et
al. 1998).  We therefore selected the longer wavelength Band~C and Band~E
images, which have isophotal central wavelengths of
$\lambda_c$=12.13$\mu$m and 21.34$\mu$m, respectively, with bandwidth of
1.72$\mu$m and 6.24$\mu$m. Consequently, $MSX$ Bands~C and E are 
narrower, higher spatial resolution analogues of the well known $IRAS$
12$\mu$m and 25$\mu$m bands. This is illustrated in Fig.~1 of Paper~I.
$MSX$ Band~D was also neglected, since its addition would not have added
to the scientific results of this work. The calibration and photometric
accuracy of $MSX$ are discussed in detail by Egan et al. (1999). The zero
magnitude flux is based on the Kurucz model for Vega (Cohen et al. 1992),
and corresponds to 9.259$\times 10^{-13}$ W m$^{-2}$ (Band~C) and
3.555$\times 10^{-13}$ W m$^{-2}$ (Band E).

Individual GH\,{\sc ii} and H\,{\sc ii} region images were obtained from
the $MSX$ Image Server at IPAC {\tt
http://irsa.ipac.caltech.edu/applications/ $MSX$/}.  For our purposes, we
sought integrated MIR fluxes in Janskys for comparison with radio fluxes.  
Consequently, the $MSX$ point source catalogue was not appropriate, given
the need to use variable apertures for these extended sources.
Extracted MIR fluxes of each GH\,{\sc ii} region are obtained from integrated 
fluxes using a circular aperture of diameter $\theta$, 
as derived from $MSX$ 21$\mu$m images (see Table 2), with a mean diameter of
$\theta$(21$\mu$m)=4 
arcmin (40 pixels). Identical aperture diameters were adopted for the 12$\mu$m flux
measurements. In all cases the apertures were centred on the radio positions, 
with the exception of a few spatially large nearby regions such as 
M17, for which an aperture of $\theta$= 9$'$, centred at G15.062--0.693 
was required to accommodate the entire GH\,{\sc ii} region.
 Thin annular
rings were taken in all cases, in order to set the background level,
taking care to avoid MIR sources in these annuli.

Spatial integration was measured in {\sc gaia} (Draper, Gray 
\& Berry 2001)  using appropriate apertures, and provides fluxes in units
of  W m$^{-2}$, after correction for the 6$''\times 6''$ pixel-area, i.e.
8.4615$\times 10^{-10}$ sr. Division by the filter bandwidth then 
provides fluxes in W m$^{-2}$ Hz$^{-1}$ (= 10$^{26}$ Jy). Finally, a 
multiplication by 1.113 is required to convert the square arc pixels into
the correct Gaussian area (Cohen 2002).  The total scale factor for
Band~C  corresponds to a multiplicative factor of 27.45 in order to
convert the IPAC integrated fluxes to mJy.

\subsection{$IRAS$ Observations}

Since the H\,{\sc ii} regions under discussion here are extended, use of
the $IRAS$ Point Source Catalog (PSC) is generally not appropriate to our
sample (e.g. as used by Codella et al. 1994). Consequently, we extracted
images for each H\,{\sc ii} region for the $IRAS$ 25$\mu$m, 60$\mu$m and
100$\mu$m filters using  the $IRAS$ Sky Survey Atlas (ISSA) server, also
at IPAC  {\tt http://irsa.ipac.caltech.edu/applications/ ISSA/}. In
contrast with $MSX$, $IRAS$ was not designed to achieve absolute surface
brightnesses. Nevertheless, ISSA was optimized to permit flux extraction
for ($\sim$arc minute sized) extended  sources (Gautier 2003). For
comparison, Chan  \& Fich (1995) used earlier Full REsolution Survey
CO-adder (FRESCO) data products for their study of H\,{\sc ii} regions
in the range 70$<l<235$.

Again, {\sc gaia} was used to spatially integrate the ISSA images.  The
apertures were generally centred on the radio peak with apertures selected
to measure the entire flux associated with the GH\,{\sc ii} region in the
60$\mu$m filter, whilst attempting
to neglect the contribution from neighbouring regions.  Identical apertures
were adopted for other filters. Since ISSA pixels are uniformly 1.5 arcmin
across, $ISSA$ apertures were correspondingly larger than $MSX$ apertures,
with a mean diameter of $\theta$(60$\mu$m)=10.5 arcmin (7 pixels).
ISSA images provide fluxes in
MJy\,sr$^{-1}$. The solid angle of one ISSA pixel is (1.5/(57.3 $\times$
60))$^2$ = 1.90386 $\times 10^{-7}$, such that one needs merely to
multiply spatially summed counts by a factor of 0.190386 to obtain
integrated fluxes in Jy. We have compared ISSA
25$\mu$m to MSX 21$\mu$m fluxes in all cases, 
revealing an average ratio of 1.7$\pm$0.8. One would expect the IRAS
fluxes to be greater owing to the greater bandpass and lower spatial resolution.

\begin{small}
\begin{table*}
\caption{ $MSX$ 12$\mu$m (Band~C) and 21$\mu$m (Band E) and 
$IRAS$ ISSA (60$\mu$m and 100$\mu$m) fluxes for radio selected Galactic 
GH\,{\sc ii} regions, including a brief summary of the 21$\mu$m spatial
morphology, namely single (s) vs multiple (m), compact (c) vs elongated (e),
containing diffuse emission (d) or not.}
\begin{tabular}{l@{\hspace{2.5mm}}
r@{\hspace{1.5mm}}
r@{\hspace{3mm}}
r@{\hspace{3mm}}
r@{\hspace{1.5mm}}
r@{\hspace{1.5mm}}
r@{\hspace{1.5mm}}
r@{\hspace{1.5mm}}
r@{\hspace{1.5mm}}
r@{\hspace{1.5mm}}
r@{\hspace{1.5mm}}
r@{\hspace{1.5mm}}
r@{\hspace{3mm}}
l}
\hline
Name & $l$ & $b$ & $d$ & log F(Jy) & log F(Jy) & $\theta$(arcmin) & log F(Jy) 
&log F(Jy)& log F(Jy) &  $\theta$(arcmin)& log L(IR) & log N(LyC)&Morph\\
     &     &     & kpc & 12$\mu$m & 21$\mu$m & 21$\mu$m & 25$\mu$m& 60$\mu$m & 100$\mu$m & 60$\mu$m& $L_{\odot}$& s$^{-1}$& \\
\hline
& 0.361 & -0.780 & 8.0 & 1.82 & 2.56 & 4.4 & 2.63 & 3.34 & 3.09 & 8.4 & 5.49 & 50.02 & m, 
e, d\\
& 0.394 & -0.540 & 8.0 & 2.14 & 2.93 & 8.0 & 2.87 & 3.62 & 3.34 & 9.0 & 5.79 & 50.03 & m, 
c, d\\
& 0.489 & -0.668 & 8.0 & 1.31 & 2.11 & 3.6 & 2.23 & 3.43 & 3.78 & 8.0 & 5.60 & 50.12 & m, 
c, d\\
W24& 0.518 & -0.065 & 8.0 & 2.82 & 3.55 & 6.4 & 3.58 & 4.63 & 4.72 & 10.5 & 6.76 & 50.91 & m, c, d\\
& 0.572 & -0.628 & 8.0 & 2.17 & 2.82 & 4.8 & 2.96 & 3.66 & 3.80 & 9.0 & 5.84 & 50.05 & s, e, d\\
Sgr~D& 1.149 & -0.062 & 8.0 & 2.55 & 3.01 & 6.6 & 3.10 & 3.94 & 4.30 & 
10.4 & 6.16 & 50.52 & s, c, d\\
& 2.303 & 0.243 & 14.3 & 1.60 & 2.28 & 4.0 & 2.61 & 3.44 & 3.53 & 10.0 & 6.06 & 50.45 & s, e, d\\
AMWW 34& 3.270 & -0.101 & 14.3 & 1.63 & 2.31 & 4.0 & 2.78 & 3.78 & 3.95 & 10.0 & 6.39 & 50.47 & s, e, d\\
AMWW 35& 4.412 & 0.118 & 14.6 & 1.65 & 2.28 & 2.4 & 2.75 & 3.67 & 3.96 & 10.0 & 6.34 & 50.56 & s, e, d\\
M8& 5.973 & -1.178 & 2.8 & 3.05 & 3.69 & 4.0 & 4.11 & 4.62 & 4.55 & 13.0 & 5.84 & 50.19 & s, c, d\\
& 8.137 & 0.228 & 13.5 & 2.26 & 2.80 & 2.2 & 2.93 & 3.79 & 3.95 & 9.0 & 6.40 & 50.47 & s, c\\
W31& 10.159 & -0.349 & 4.5 & 3.23 & 3.75 & 4.8 & 3.77 & 4.52 & 4.62 & 9.0 & 6.22 & 50.66 & m, c\\
W31& 10.315 & -0.150 & 15.0 & 2.75 & 3.37 & 3.4 & 3.35 & 4.06 & 4.17 & 8.0 & 6.82 & 50.90 & m, c\\
M17& 15.032 & -0.687 & 2.4 & 4.35 & 4.95 & 9.0 & 4.92 & 5.33 & 5.35 & 20.0 & 6.61 & 51.22 & m, e, d\\
& 20.733 & -0.087 & 11.8 & 2.28 & 2.92 & 4.8 & 3.25 & 4.15 & 4.30 & 14.0 & 6.61 & 50.72 & m, c, d\\
W42& 25.382 & -0.177 & 11.5 & 3.00 & 3.63 & 4.6 & 3.64 & 4.32 & 4.34 & 10.0 & 6.84 & 50.93 & m, c \\
   & 29.944 & -0.042 & 6.2 & 2.79 & 3.39 & 5.0 & 3.33 & 4.18 & 4.28 & 10.0 & 6.15 & 50.33 & m, c\\
W43& 30.776 & -0.029 & 6.2 & 3.19 & 3.71 & 4.0 & 3.82 & 4.61 & 4.72 & 10.0 & 6.56 & 50.83 & m, e\\
AMWW 52& 32.797 & 0.192 & 12.9 & 1.66 & 2.28 & 1.4 & 2.42 & 3.54 & 3.72 & 9.0 & 6.09 & 50.03 & s, c\\
W49A& 43.169 & 0.002 & 11.8 & 2.90 & 3.47 & 4.0 & 3.59 & 4.55 & 4.74 & 10.0 & 7.04 & 51.21 & m, c, d\\
   & 48.596 & 0.042 & 9.8 & 2.22 & 2.63 & 4.0 & 2.85 & 3.93 & 4.11 & 12.0 & 6.23 & 50.14 & m, c, d\\
W51& 48.930 & -0.286 & 5.5 & 2.91 & 3.54 & 5.4 & 3.66 & 4.23 & 4.18 & 8.0 & 6.10 & 50.03 & m, e, d\\
W51West& 49.384 & -0.298 & 5.5 & 2.72 & 3.27 & 4.8 & 3.34 & 4.23 & 4.43 & 8.0 & 6.08 & 50.15 & m, c\\
W51A& 49.486 & -0.381 & 5.5 & 3.29 & 3.94 & 4.4 & 4.16 & 4.84 & 4.97 & 13.0 & 6.68 & 50.94 & m, c\\
W58A& 70.300 & 1.600 & 8.6 & 2.78 & 3.31 & 2.5 & 3.35 & 4.21 & 4.27 & 11.0 
& 6.43 & 50.06 & m, c\\
DR7& 79.293 & 1.296 & 8.3 & 2.16 & 2.73 & 2.5 & 3.05 & 3.90 & 3.97 & 12.0 
& 6.05 & 50.15 & m, e\\
W3& 133.720 & 1.210 & 4.2 & 3.25 & 3.98 & 4.0 & 4.06 & 4.70 & 4.84 & 12.0 & 6.35 & 50.25 & m, c\\
RCW42& 274.013 & -1.141 & 6.4 & 2.81 & 3.42 & 3.2 & 3.52 & 4.30 & 4.35 & 12.0 & 6.27 & 50.36 & m, e\\
& 282.023 & -1.180 & 5.9 & 2.73 & 3.30 & 3.0 & 3.43 & 4.17 & 4.18 & 10.0 & 6.06 & 50.32 & s, c\\
RCW49& 284.301 & -0.344 & 4.7 & 3.75 & 4.37 & 9.2 & 4.43 & 5.09 & 5.08 & 22.0 & 6.82 & 50.96 & s, e, d\\
NGC 3372& 287.379 & -0.629 & 2.5 & 3.38 & 4.12 & 9.0 & 4.49 & 5.00 & 5.03 & 18.0 & 6.14 & 50.11 & m, e, d\\
& 289.066 & -0.357 & 7.9 & 1.97 & 2.42 & 3.4 & 2.83 & 3.82 & 4.04 & 13.0 & 5.93 & 50.05 & s, e, d\\
NGC 3576& 291.284 & -0.712 & 3.1 & 3.28 & 4.05 & 3.0 & 4.03 & 4.64 & 4.80 & 12.0 & 6.07 & 50.28 & s, c\\
NGC 3603& 291.610 & -0.528 & 7.9 & 3.88 & 4.55 & 7.4 & 4.59 & 5.06 & 5.07 & 18.0 & 7.31 & 51.50 & m, c, d\\
& 298.227 & -0.340 & 10.4 & 3.00 & 3.69 & 2.8 & 3.74 & 4.30 & 4.28 & 10.4 & 6.75 & 50.87 & s, c\\
& 298.862 & -0.438 & 10.4 & 2.73 & 3.39 & 4.6 & 3.63 & 4.42 & 4.46 & 15.0 & 6.78 & 50.87 & m, c, d\\
& 305.359 & 0.194 & 3.5 & 2.64 & 3.17 & 2.4 & 3.52 & 4.35 & 4.40 & 8.0 & 5.75 & 50.13 & m, c\\
& 319.158 & -0.398 & 11.5 & 1.85 & 2.39 & 3.5 & 2.71 & 3.67 & 3.86 & 9.0 & 6.12 & 50.30 & s, c, d\\
& 319.392 & -0.009 & 11.5 & 2.26 & 2.93 & 5.2 & 3.00 & 3.87 & 3.96 & 10.4 & 6.34 & 50.17 & m, c, d\\
& 320.327 & -0.184 & 12.6 & 2.27 & 2.72 & 3.6 & 2.83 & 3.75 & 3.95 & 9.0 & 6.31 & 50.11 & m, c\\
RCW97& 327.304 & -0.552 & 3.0 & 3.05 & 3.65 & 3.4 & 3.82 & 4.60 & 4.79 & 12.0 & 5.92 & 50.14 & m, c \\
& 327.993 & -0.100 & 11.4 & 2.12 & 2.48 & 3.1 & 2.64 & 3.59 & 3.71 & 9.0 & 6.04 & 50.08 &s, c, d\\
& 330.868 & -0.365 & 10.8 & 2.12 & 2.62 & 2.6 & 3.08 & 4.12 & 4.30 & 9.0 & 6.49 & 50.56 & m, c, d\\
& 331.324 & -0.348 & 10.8 & 2.15 & 2.71 & 2.3 & 2.90 & 3.67 & 3.81 & 8.0 & 6.09 & 50.28 & s, c, d\\
& 331.354 & 1.072 & 10.8 & 2.12 & 2.66 & 2.3 & 2.86 & 3.69 & 3.84 & 8.0 & 6.11 & 50.10 & s, c \\
& 331.529 & -0.084 & 10.8 & 2.92 & 3.44 & 4.4 & 3.52 & 4.48 & 4.61 & 9.0 & 6.89 & 51.16 & m, e, d\\
& 333.122 & -0.446 & 3.5 & 2.78 & 3.38 & 3.0 & 3.64 & 4.52 & 4.66 & 8.0 & 5.93 & 50.08 & s, c\\
& 333.293 & -0.382 & 3.5 & 2.86 & 3.50 & 4.0 & 3.60 & 4.39 & 4.54 & 9.0 & 5.84 & 50.04 & m, c\\
& 333.610 & -0.217 & 3.1 & 3.65 & 4.10 & 3.6 & 4.12 & 4.55 & 4.60 & 9.0 & 6.03 & 50.43 &s, c\\
& 338.398 & 0.164 & 13.1 & 1.93 & 2.65 & 2.5 & 3.45 & 4.23 & 4.27 & 9.0 & 6.73 & 50.90 & m, e, d\\
& 338.400 & -0.201 & 15.7 & 1.28 & 1.87 & 1.4 & 2.13 & 3.30 & 3.48 & 8.0 & 6.00 & 50.24 & s, c \\
& 345.555 & -0.043 & 15.2 & 1.96 & 2.54 & 2.4 & 2.77 & 3.74 & 3.90 & 8.0 & 6.43 & 50.69 & m, c, d\\
& 345.645 & 0.009 & 15.2 & 1.56 & 2.17 & 1.9 & 2.49 & 3.64 & 3.75 & 8.0 & 6.30 & 50.53 & s, c\\
& 347.611 & 0.204 & 6.6 & 2.16 & 2.91 & 2.6 & 3.38 & 4.17 & 4.28 & 9.0 & 6.11 & 50.44 & m, e, d\\
& 351.467 & -0.462 & 13.7 & 2.07 & 2.43 & 2.2 & 2.69 & 3.65 & 3.81 & 9.0 & 6.25 & 50.17 &s, c\\
Sgr~C& 359.429 & -0.090 & 8.0 & 2.16 & 2.93 & 3.0 & 3.30 & 4.53 & 4.66 & 9.0 & 6.63 & 50.46 & s, c, d\\
\hline
30 Dor& 279.448 & -31.678 & 50.0 & 2.50 & 3.21 & 4.4 & 3.32 & 3.98 & 3.93 & 12.0 & 7.74 & 52.32&s, e, d\\
NGC 346& 302.100 & -44.930 & 60.0 & 0.35 & 1.23 & 2.8 & 1.51 & 2.43 & 2.48 & 10.0 & 6.29 & 51.05 & s, e, d\\
\hline
\hline
\end{tabular}  
%(a) Downes et al. (1980); (b) Israel (1980); (c) Araya et al. (2002); (d) Watson et al. (2003); (e) Sewilo et al. (2004)
\end{table*}
\end{small}

\section{MIR MOPHOLOGIES}\label{morph}

In the Appendix, available electronically, Figs. A1--10 of $MSX$ Band~E (21$\mu$m) images are
presented of all Galactic and Magellanic Cloud GH\,{\sc ii} regions. 
Each of these figures contains six objects, with a field 
of view of 15$\times 15$ arcmin.  The intensity is in logarithmic units 
(W m$^{-2}$ sr$^{-1}$), scaled between the brightest knots and those 1.0
dex fainter, as indicated in the adjacent key. In these figures one is 
looking at the integrated dust emission, which comes from unresolved
sources, resolved objects (the cluster dust), and from an extended halo
associated with the formation process. In Table 2, we indicate whether individual
sources are single or multiple, compact or elongated, or are associated with
diffuse components.

All of the sources are resolved with $MSX$, that is, their sizes 
are greater than 18$''$, the spatial resolution of the instrument.  This
corresponds to different physical scales, depending on the source
distances, as may be seen in the images.  Many of them have multiple
components, suggesting several star forming clusters are present within 
the ionized hydrogen regions (e.g. W49A).  The sources in this paper 
frequently show diffuse emission surrounding them, often with loops and
tails. This is in contrast to the single star excited UCH\,{\sc ii}
regions of Paper~I where such features were less common. Unattributed
comments concerning  NIR photometry of individual sources are from
unpublished observations (Blum, private communication). In many cases,
the MIR $MSX$ images closely mimic radio continuum datasets, also
common to UCH\,{\sc ii} regions from Paper~I, and the study of Cohen \&
Green (2001).

We now discuss a subset of the GH\,{\sc ii} regions, grouped according to
whether they are observed optically or in the NIR.

\subsection{Optically visible GH\,{\sc ii} regions}

Within the Milky Way, 
$\sim$15\% of the GH\,{\sc ii} regions are optically visible --
namely NGC~3372 (Carina), M8, M17 (Omega), RCW42, RCW49, RCW97, 
NGC~3576 and NGC~3603. 
Their MIR morphologies are complex, ranging from a 
peculiar V-shape of the dust in the case of M17, which contains a luminous
cluster (Hanson et al. 1997), to several
distinct regions of very bright dust emission in Carina (Rathborne et al. 2004)
and NGC~3603 (N\"{u}rnberger 2002). 

Rathborne et al (2004) present a detailed
morphological discussion of Carina, such that it is not discussed here, except
that the measurements presented here are centred on the 
Northern Molecular 
Cloud, containing the cluster Tr~14, such that $\eta$ Carina is excluded. 
For NGC~3603, both the brighter, compact north-west source and
extended SW source are included in our $MSX$ measurements.
The optically visible cluster 
lies between these two IR sources slightly to the north, the stellar content 
of which is well studied (e.g. Moffat 1983), and contains three 
H-rich Wolf-Rayet stars (Drissen et al. 1995). Finally,
a NIR image of the star  cluster associated with NGC~3576 is given 
by Figuer\^{e}do et al. (2002), who also include K band spectroscopy.  

 A
strong elongated MIR source is found at the radio position of RCW49, with
a similar morphology to M17. A cluster is present in NIR images, whilst
the Westerlund~2 cluster lies immediately to the north of the peak MIR    
emission, which hosts two WR stars (Shara et al. 1991). Finally, RCW42
reveals a bright MIR    source, with an extended halo, which is
spatially coincident  with a NIR cluster.

Turning to extragalactic sources,
30 Doradus, alias NGC~2070 (Tarantula Nebula), is the prototype optically
visible GH\,{\sc ii} region of the Local Group.
Spectroscopy of stars within the central cluster, R136, reveals a large
concentration of early O stars (Massey \& Hunter 1998), which also hosts 
a number of Wolf-Rayet stars (e.g. Walborn 1973; Crowther \& Dessart
1998). In the $MSX$ images, dust emission appears as an incomplete torus
with an opening to the east. NGC~346 is rather less massive, although it hosts
many young OB stars (Massey et al. 1989). Although fairly weak,
NGC~346 is spatially extended NW to SE in the $MSX$ image. 

\subsection{NIR GH\,{\sc ii} regions}

Approximately 2/3 of the GH\,{\sc ii} regions can be 
readily identified on K-band 2MASS images via diffuse nebulosity. 
$MSX$ images of
W3 have already been presented by Kraemer et al. (2003), and so is not
discussed further here. In addition, several sources host 
UCH\,{\sc ii} regions, including  G29.944--0.042, 
G70.300+1.600 (K3--50A) and G79.293+1.296 (DR7), as discussed in Paper~I,
the latter hosting a NIR cluster (Dutra \& Bica 2001).

G10.159--0.349 (W31), W42 and G333.610--0.217 
have a similar MIR spatial appearance, with a compact
spherical core surrounded by an extended halo. Each  
host a luminous cluster as revealed by NIR images of 
Blum et al. (2000, 2001, 2003). In contrast,
W43 includes three separate  dust emission sources, 
of which the central source contains a young cluster, itself hosting 
a Wolf-Rayet star at its core. W49A is more complicate still, revealing 
several distinct distinct dust emission sources that 
correspond to individual clusters of OB stars (Conti \& Blum 2002). Note 
the MIR spatial morphology  of W49A is exactly
matched by Br$\gamma$~ emission on the deep NIR image of this field by
Alves \& Homeier (2003). GH\,{\sc ii} regions within the W51 complex exhibit
similar spatial morphologies.

G298.227--0.340 and G351.467--0.462 have a rather simple MIR morphology, with 
compact cores lacking an extended halo.  These source, alias IRAS~12073--6233
and IRAS~17221--3619,  were included in the IR/radio study of H\,{\sc ii} regions by
Martin-Hernandez et al. (2003).
Finally, the compact NIR H\,{\sc ii} region identified in G1.149--0.062 by
Blum \& Damineli (1999) is spatially coincident with the dominant MIR source
Sgr~D at G1.128--0.104, 3$'$ SE from the radio peak.

\subsection{MIR GH\,{\sc ii} regions}

The remanding $\sim$1/3 of the GH\,{\sc ii} region sample exhibit 
neither optical nor NIR signatures associated with the radio/MIR source. 
For this subsample, the extinction to the ionizing source is either rather high or 
in a few cases identification as GH\,{\sc ii} regions may be in question. 
MIR morphologies
for this group span a wide range. In some cases the GH\,{\sc ii} regions
reveal a compact dust source with associated extended
halo (e.g. G345.555--0.043) that resemble the morphology of 
regions with luminous NIR clusters such as W31. In other cases
bright, extended dust sources are observed (e.g. G2.303+0.243), 
reminiscent of RCW49. Several exhibit morphologies involving 
multiple point sources plus diffuse emission (e.g. W24, 
G20.733--0.087). Finally, several show primarily diffuse dust emission 
(e.g. G0.489--0.668, G3.270--0.101).

\subsection{Colour-colour diagrams}

The dust emission of the GH\,{\sc ii} regions arises
primarily from the interstellar material within or close to the exciting
star clusters.  Some emission may also have an origin from individual
circumstellar dust cocoons surrounding UCH\,{\sc ii} regions (or hot 
cores) within the clusters.  As with stellar magnitudes UBV or JHK, it
will be instructive to plot analogous colors for the MIR or FIR fluxes. 
In Figure 2(a) we make IR colour-colour plots using the log
flux ratio (colour) 100$\mu$m/60$\mu$m versus 60$\mu$m/25$\mu$m for the
UCH\,{\sc ii} of Paper I (open) and for the GH\,{\sc ii} regions of this paper
(filled). Similar plots for a smaller number of normal H\,{\sc ii} regions
were presented by Chan \& Fich (1995).

We see immediately that the sources in the two plots have very similar
colours, although there is more scatter in the UCH\,{\sc ii} points.
Recall that Wood \& Churchwell (1989) used colour--colour plots
similar to these to identify UCH\,{\sc ii} regions. We see that such
criteria will also pick out GH\,{\sc ii} regions. The outliers
in Fig 2(a) with the lowest  100$\mu$m/60$\mu$m ratios are G0.361--0.780 
and
G0.394--0.540.

To get a feeling for the dust temperatures associated with these colours,
we used the black body distributions convolved with the IRAS filter
profiles (based on Beichman et al. 1988). In each plot there is a main
body of points with outliers, the most extreme of which are usually
those with complicated morphology. In both cases we see that on average
the 60/25$\mu$m colours lead to higher temperatures than the 
100/60$\mu$m colours.  This is a quantitative way to demonstrate that the
SED of the dust of UCH\,{\sc ii}, GH\,{\sc ii} and H\,{\sc ii}'s do not
follow simple black body laws. Note that the main body of points of the
UCH\,{\sc ii} are at slightly cooler temperatures but there is plenty of
overlap. 

Using the MSX data we have also made an analogous plot of 21/12$\mu$m
versus the IRAS 60/25$\mu$m colors, which is presented in Fig.~2(b). The
inferred temperatures were still higher with these shorter wavelength $MSX$
colours, although we acknowledge that the 12$\mu$m fluxes are influenced
by the silicate absorption feature which affects this wavelength. The observed
scatter for the complex morphology UCH\,{\sc ii} regions (open triangles) 
was much greater than for the simple UCH\,{\sc ii} regions (open squares) and
GH\,{\sc ii} regions (filled circles).

\begin{figure} 
\epsfxsize=8.8cm\epsfbox[20 25 524 775]{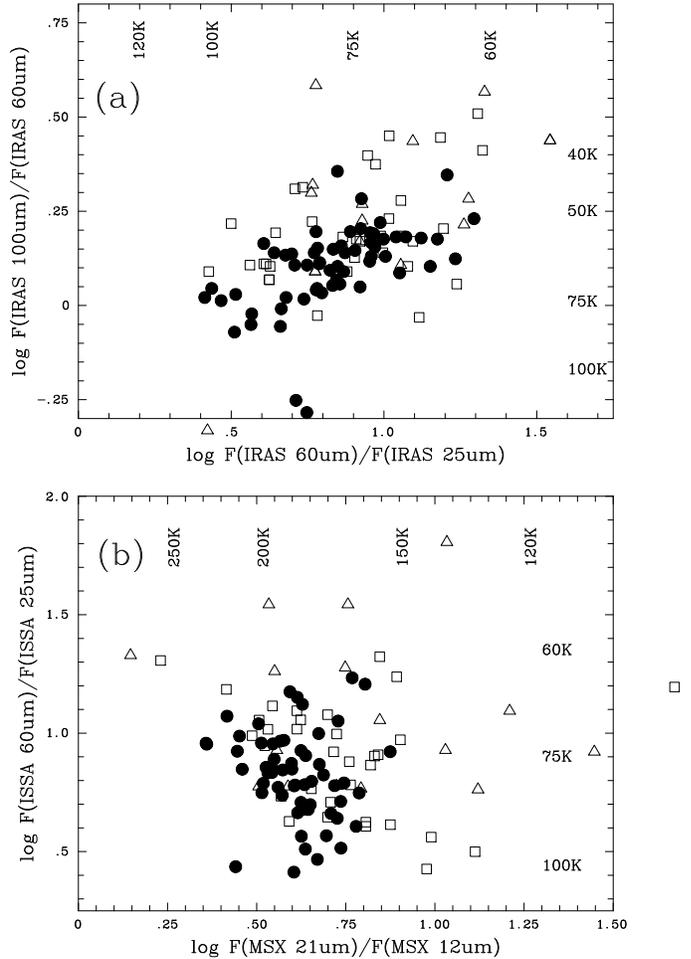} 
\caption{Colour-colour diagrams: {\bf (a)}: $IRAS$ log flux ratio (colour)
100$\mu$m/60$\mu$m versus 60$\mu$m/25$\mu$m for the GH\,{\sc ii} regions
newly presented here (filled circles) and UCH\,{\sc ii} regions from Paper~I
(open squares: simple morphology, open triangles: complex morphology).
 {\bf (b):} log flux ratio (colour) of $IRAS$ 60$\mu$m/25$\mu$m versus
$MSX$ 21$\mu$m/12$\mu$m for the GH\,{\sc ii} regions (filled circles).}
\label{GHII_CCD}
\end{figure}

\section{RELATION BETWEEN RADIO EMISSION AND FIR FLUXES}\label{flux}

The radio emission of H\,{\sc ii} regions arises from the integrated
Lyman continuum luminosity (EUV) of the hot star content of the
cluster(s) within. What is measured is the free-free emission of the
ionized hydrogen and/or the strength of high level recombination
hydrogen emission lines, typically obtained at $cm$ wavelengths. These 
can be converted to the number of Lyman continuum photons (NLyC)
producing the emission using well known physical relationships, {\em if
the distance is known}.  What one wishes to obtain is the NLyC photons
emitted by the stars, so that one may infer their stellar properties.
The observed NLyC could  be affected by leakage of photons from the
region without ionizing the material, a so-called density bounded
nebulosity.  Some emitted photons might also be absorbed by the dust
within the nebulae.  In both cases, the {\em observed} NLyC would be
smaller than the {\em emitted} NLyC and one would underestimate the hot
star content.
How important are these effects for GH\,{\sc ii} regions?

UCH\,{\sc ii} regions have very dense cocoons overlaying the ionized
hydrogen but also tend to have extended hydrogen emission halos,
suggesting that their inner environments are density bounded and/or
clumping may play a role (Ignace \& Churchwell 2003) and LyC photons are
escaping. Whether clusters are density bounded is a longstanding problem
that is not fully resolved, but does not need to be settled here. It
would be important to resolve this issue were we to try to infer the
stellar content of our clusters from the counted NLyC photons alone.

We now turn to the role of the dust.  Its emission is related to the
quantity and quality of material present, and the heating by 
the hot stars in the clusters.  Their emitted radiation
peaks in the far UV beyond the Lyman limit; the thermal dust emission 
peaks near 100$\mu$m.  We
have estimated the total IR integrated luminosity (L$_{\rm IR}$) of the dust
using a standard relation (e.g. Chan \& Fich 1995), with a minor
modification to include the more reliable $MSX$ MIR fluxes, via:
\begin{eqnarray} 
 L_{\rm IR} &\sim &1.58 (d/{\rm kpc})^2 \left[1.3(F_{12}^{\rm MSX} + F_{21}^{\rm
MSX}) \nonumber \right.\\
& & \left.+ 0.7(F_{21}^{\rm MSX} + F_{60}^{\rm ISSA}) + 0.2(F_{60}^{\rm
ISSA} + F_{100}^{\rm ISSA}) \right] \nonumber
\end{eqnarray} 
If the dust is optically thick to stellar radiation, the L$_{\rm IR}$ is a good
estimate of the L$_{bol}$.  With optically thin dust, stellar
radiation escapes to space and is not available to heat the material
and the LIR will underestimate the total luminosity L$_{bol}$.

In Fig.~3, we plot the observed L$_{\rm IR}$ inferred from $MSX$ and $ISSA$,
versus log N(LyC), which are both distant dependent. This {\em ratio} is
distance independent but strongly spectral type dependent (e.g.,
Churchwell 2002). (In Paper I, we made an analogous plot but used the
100$\mu$m luminosity for the ordinate. That relationship is similar to
that shown here.) The UCH\,{\sc ii} are to the right at lower log
N(LyC); the GH\,{\sc ii} to the left. H\,{\sc ii} regions would fill the 
space immediately  to the right of the GH\,{\sc ii}. 
%Comparing the two
%groups one sees that the cluster sources are on average below
%those of the single star excited UCH\,{\sc ii} objects  projected
%towards the same N(LyC).  
%
The lines represent model predictions from
Lumsden et al. (2003) for clusters with a nominal Salpeter
IMF (solid line: upper) and single stars (dotted line: lower) based on
CoStar stellar models of Schaerer et al. (1996). Similar results regarding
Lyman continuum ionizing photons are obtained with alternative WM-basic
models (Smith et al. 2002). Whilst the majority of the 
UCH\,{\sc ii} regions lie between
the cluster and single star predictions, the GH\,{\sc ii} regions actually
lie below the (extrapolated) single star predictions -- their
IR luminosity is a factor of a few too low versus the cluster predictions. 
What is going on here? We assert that the problem
is that the models assume {\em all} the stellar radiation is absorbed by
the dust, which is optically thick, thus L$_{\rm IR}$ = L$_{bol}$. It could be
that some of the UCH\,{\sc ii} regions, which lie close to, or just
above, their model predictions, are optically thick.  But the clusters
are {\em not}.

What are we to make of the large number of UCH\,{\sc ii} regions between
the single star and cluster predictions? The most reasonable explanation
is that these objects contain extra stars which contribute to the L$_{\rm IR}$
but not so much to the N(LyC) numbers. They behave, therefore, as
miniclusters. This explanation was also given by Lumsden et al. (2003). 
But there are
still a dozen or so UCH\,{\sc ii} above the cluster predictions. This
cannot be an optical depth effect as smaller values would push the L$_{\rm IR}$
lower.  It is possible that some of these sources have either dust
absorption of the LyC photons and/or they are density bounded thus they 
would be moved to the left. Indeed,
even some of the UCH\,{\sc ii} regions located between the model
predictions could have these problems.  Note that a couple of UCH\,{\sc
ii} regions are below the single star predictions.  These might not
be optically thick. 

Let us now turn to understanding the placement of the GH\,{\sc ii}
regions in Fig 3.
There is, curiously, less scatter in the vertical dispersion than for
the UCH\,{\sc ii} regions (one would have expected more scatter given
the larger number of free parameters for the clusters). We can assert
that those clusters along the top edge of the relationship are on average
dustier, or have a higher optical depth, than those along the lower
boundary. The vertical or horizontal scatter is a factor of a
few. While we may be miscounting the N(LyC) emitted by the stars which
would move them to the left, we suspect that the optical depth effect in
the dust is the more important variable here.\footnote{Errors in the
distance affect both axes similarly.} 
Included among the lower boundary group is NGC~346, 
which contains evolved O
type stars, thus is somewhat older than the others.  The vertical range
of the clusters of Fig.~3 at a given N(LyC) could then be mostly an age 
effect as the dust is gradually
being dissipated and the optical depth decreasing. The upper edge of the
cluster distribution is a factor of 2 or so below the model predictions.
This gives an estimate of the overall dust optical depth to stellar 
radiation, which would be of order $\tau$=0.7.

\begin{figure} 
\epsfxsize=8.8cm\epsfbox[0 400 504 790]{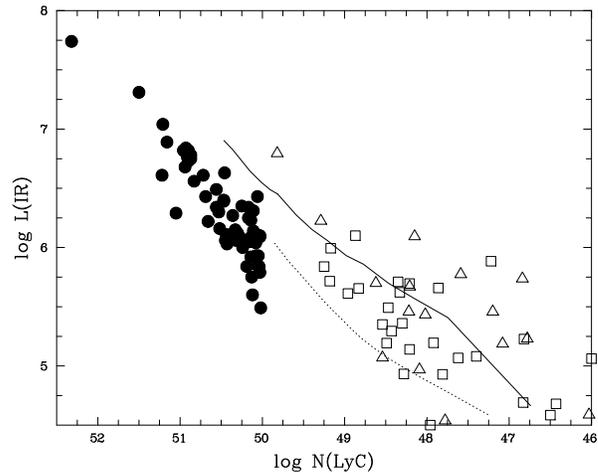}
\caption{Comparison between Lyman continuum log flux inferred from radio
free-free emission and IR luminosity of UCH\,{\sc ii} (from Paper~I,
open symbols) plus Galactic and Magellanic Cloud GHII\,{\sc ii} regions
(filled symbols) inferred from $ISSA$ and $MSX$. The solid lines are
the predictions of the CoStar
models from Lumsden et al. (2003) for clusters,
whilst  dotted line are for single stars.}
\label{GHII-100um-NLyC}
\end{figure}
 
\section{SUMMARY}\label{summ}

We have carried out a complete census of 6cm selected H\,{\sc ii}
regions, for which 56  giant H\,{\sc ii} regions are identified, based in
part on  mid and far-IR fluxes from $MSX$ and $IRAS$.
Of course, the radio selected GH\,{\sc ii} catalogue presented here
is not complete. In addition, there are several clusters close to the
centre of the Galaxy with ionizing fluxes in excess of 10$^{50}$ s$^{-1}$.
These include the Arches (Serabyn et al. 1998) and 
Quintuplet (Figer et al. 1999)
clusters. In addition, older massive clusters that have already lost the
bulk of their associated gas are also omitted (e.g. Westerlund 1:
Clark \& Negueruela 2002).

 We present 
21$\mu$m images of all giant H\,{\sc ii} regions using $MSX$.
% 21$\mu$m with a spatial resolution of 18$''$. 
This wavelength
is sampling the thermal emission from the dust in the vicinity  of these
OB clusters, material left over from the birth of these objects.
The spatial morphology of most of the dust emission is complex; many
sources are double or multiple, suggesting several common sites of recent
star formation. Extended diffuse dust emission is also common, in
contrast to our analogous study of UCH\,{\sc ii} regions (Paper~I) where
such features were rare.  In the few cases in which an UCH\,{\sc ii}
region is found within a larger GH\,{\sc ii} region, the dust from the
former entity dominates the emission; it is only a small part of the
measured NLyC.  

We have also measured MIR and FIR emission from the $ISSA$ archive of
these same objects. We used the MSX satellite data to interpret the
FIR $IRAS$  fluxes which were obtained at a lower spatial resolution,
particularly the longest wavelength one at 100$\mu$m. These wavelengths
sample an appreciable fraction of the dust emission and can be used to
infer properties of its SED and total luminosity.   Tables 1 and 2
provide detailed information about each source, with relevant flux
measurements. A careful examination of the Tables indicates that 75\% 
of the sources show a 100$\mu$m flux larger than that at 60$\mu$m, thus the
peak of the dust emission SED for the cluster excited regions is near
the former wavelength or beyond. This was already known to be the case
for UCH\,{\sc ii} regions but is a new result for the  GH\,{\sc ii} regions.

In the colour-colour diagrams  we plot the $MSX$ and $IRAS$
flux ratios for the 
UCH\,{\sc ii} regions of Paper~I and the luminous GH\,{\sc ii}
of this paper.   There is a main body of points in each plot,
with outliers, and while the UCH\,{\sc ii} are a bit redder there is
considerable overlap in the colours with the cluster objects. Under
an over simplified assumption that each colour can be represented by a
black body temperature, we also show these numbers on the plots. The
temperatures inferred from the colour are mostly lower for the longer
wavelength colours in both cases. IR colour-colour selection criteria for 
UCH\,{\sc ii} regions will also pick up GH\,{\sc ii} and H\,{\sc ii} 
sources.

We infer the FIR luminosity by weighting and combining the IR fluxes
using the distances in the Tables; the N(LyC) is also listed
there. In Fig.~3 we show the relationship between these two measures.
With some scatter we see that as the number of OB stars (the abscissa)
increases, the dust emission (ordinate) does also (more stars). The
UCH\,{\sc ii} points lie, for the most part, between the predictions for
single stars and those for small clusters. Some objects lie well above
the cluster relationship; these most likely have some absorption or 
escape of LyC
photons from the nebula (quite possible if
there is significant clumping (Ignace \& Churchwell 2004). In both
cases the points would move to the left towards the predicted line,
sometimes substantially.  

\section*{Acknowledgements}  

We wish to thank Martin Cohen for help with $MSX$ flux calibration, Nick
Gautier for the calibration of $ISSA$ data, Tom Kuchar for providing the
H\,{\sc ii} region catalogue in digital form, and
%
% Cornelia Lang for providing 6
% cm images and fluxes of the Arches and Quintuplet regions prior to
% publication, and Sean Dougherty for radio fluxes of Westerlund~1. 
%
Bob   
Blum for providing us with NIR images of nearly all of our sources and 
invaluable advice on the manuscript. Comments by Joe Cassinelli, Ed
Churchwell and John Mathis are appreciated. PAC and PSC appreciate
continuing support by the Royal Society and the NSF, respectively.  This
research made use of data products from the Midcourse Space Experiment.  
Processing of the  data was funded by the Ballistic Missile Defense
Organization with additional support from NASA Office of Space Science.  
This research has also made use of the NASA/ IPAC Infrared Science
Archive, which is operated by the Jet Propulsion Laboratory, California
Institute of Technology, under contract with the National Aeronautics and
Space Administration. GAIA is  a Starlink derivative of the ESO Skycat
catalogue and image display tool.

%\bsp
%\label{lastpage}
%\end{document}
%\end{document}

\appendix 

\begin{figure*} {\bf Figure A1}. $MSX$ Band E images of GH\,{\sc ii}
regions. Each field in this and subsequent figures covers a field-of-view
of 15$\times$15 arcmin, and is presented in a logarithmic intensity
scale  (W\,m$^{-2}$ sr$^{-1}$).  The centre of each field refers to the
quoted coordinates of the radio peak.
\epsfysize=23cm   \epsfbox[0 -200 504 620]{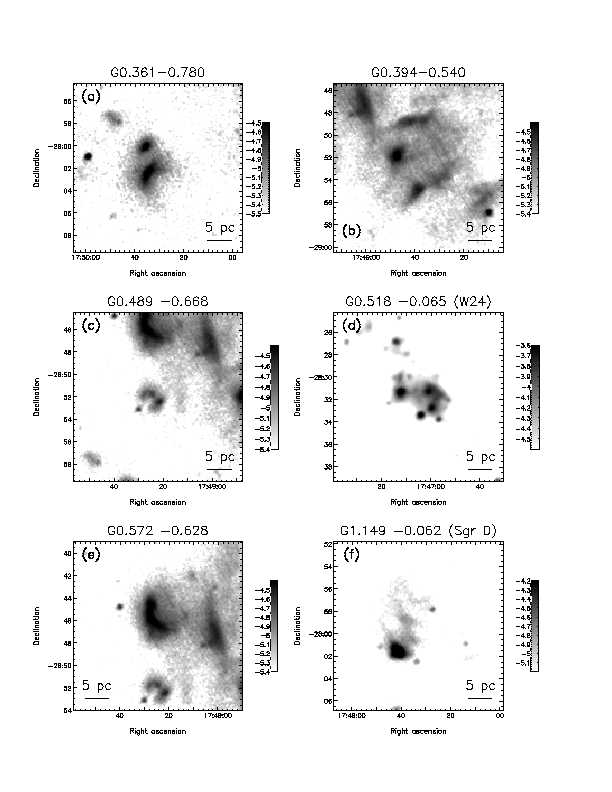}
\label{fig1}
\end{figure*}

\begin{figure*} {\bf Figure A2}. $MSX$ Band E images of GH\,{\sc ii}
regions.
\epsfysize=23cm   \epsfbox[0 -200 504 620]{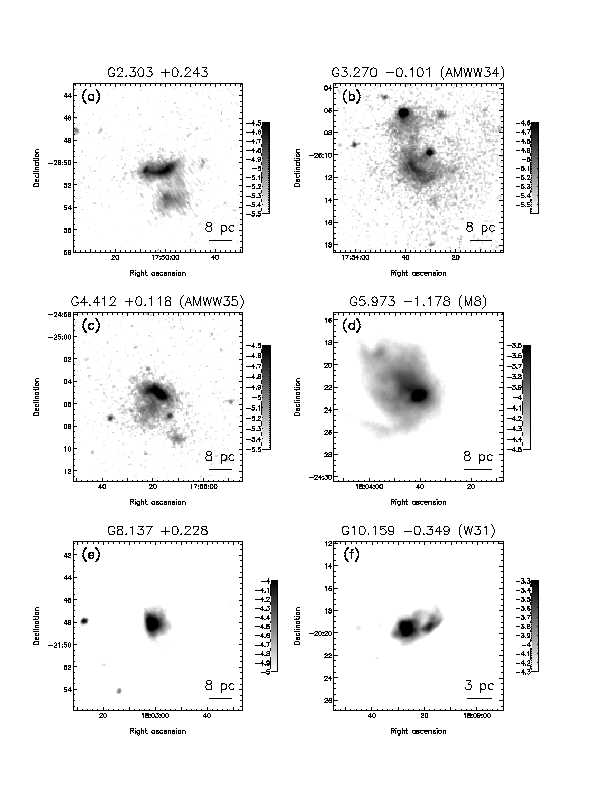}
\label{fig2}
\end{figure*}

\begin{figure*} {\bf Figure A3}. $MSX$ Band E images of GH\,{\sc ii} 
regions
\epsfysize=23cm   \epsfbox[0 -200 504 620]{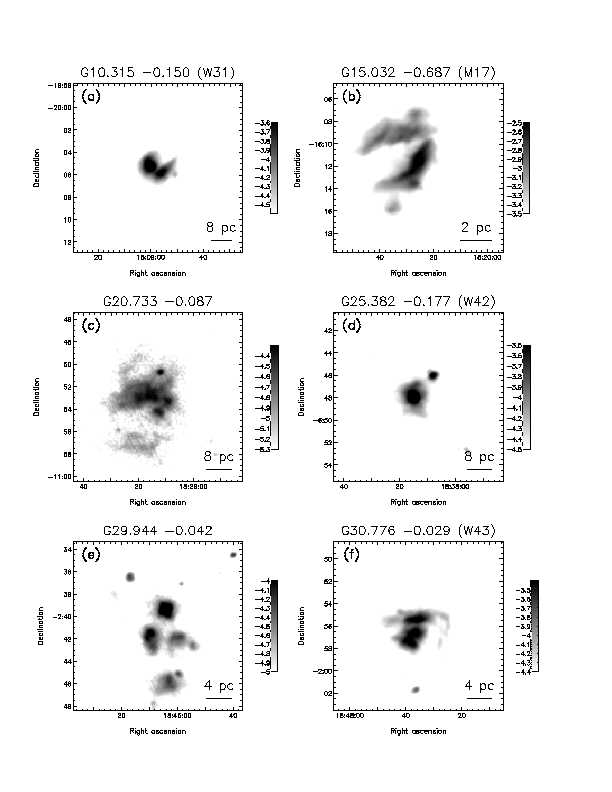}
\label{fig3}
\end{figure*}

\begin{figure*} {\bf Figure A4}. $MSX$ Band E images of GH\,{\sc ii} 
regions
\epsfysize=23cm   \epsfbox[0 -200 504 620]{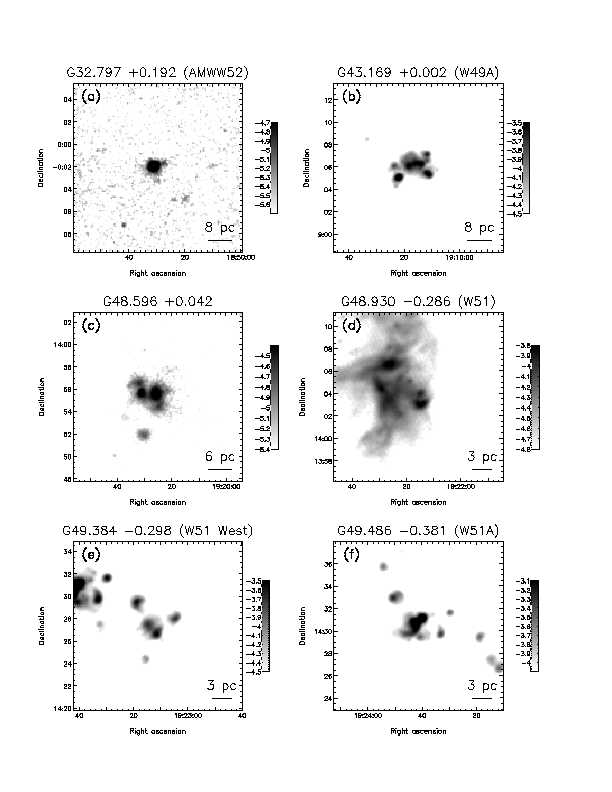}
\label{fig4}
\end{figure*}

\begin{figure*} {\bf Figure A5}. $MSX$ Band E images of GH\,{\sc ii} 
regions
\epsfysize=23cm   \epsfbox[0 -200 504 620]{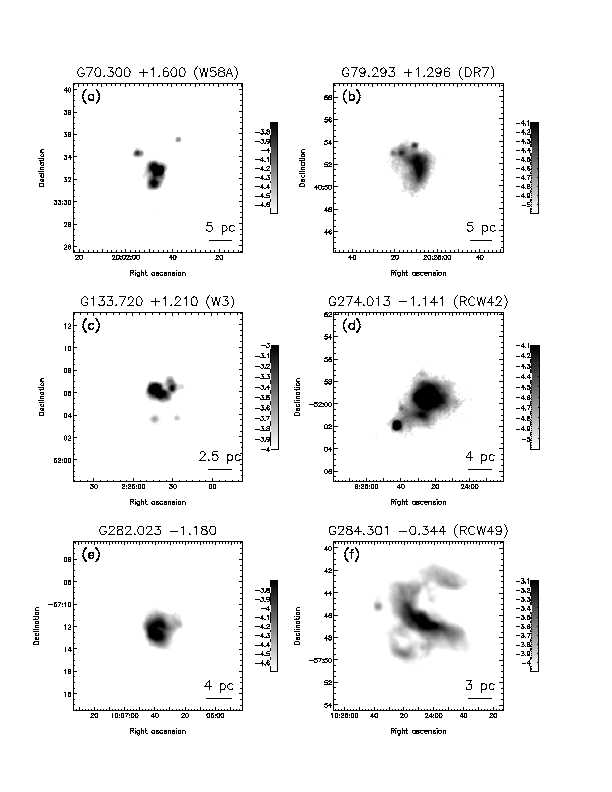}
\label{fig5}
\end{figure*}

\begin{figure*} {\bf Figure A6}. $MSX$ Band E images of GH\,{\sc ii} 
regions
\epsfysize=23cm   \epsfbox[0 -200 504 620]{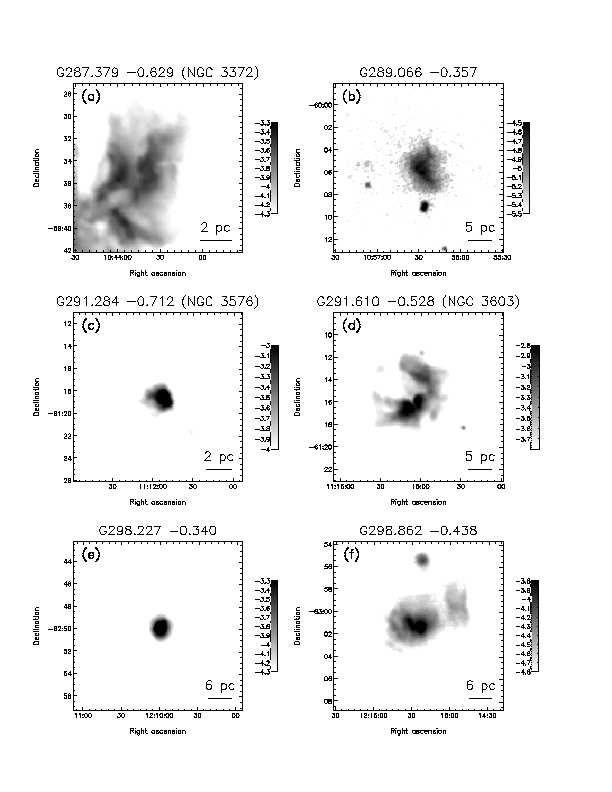}
\label{fig6}
\end{figure*}

\begin{figure*} {\bf Figure A7}. $MSX$ Band E images of GH\,{\sc ii} 
regions
\epsfysize=23cm   \epsfbox[0 -200 504 620]{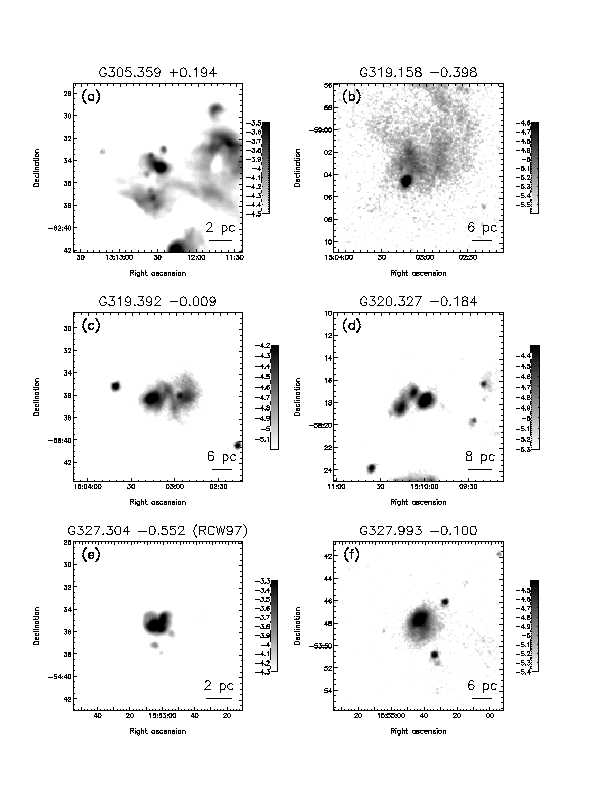}
\label{fig7}
\end{figure*}

\begin{figure*} {\bf Figure A8}. $MSX$ Band E images of GH\,{\sc ii} 
regions
\epsfysize=23cm   \epsfbox[0 -200 504 620]{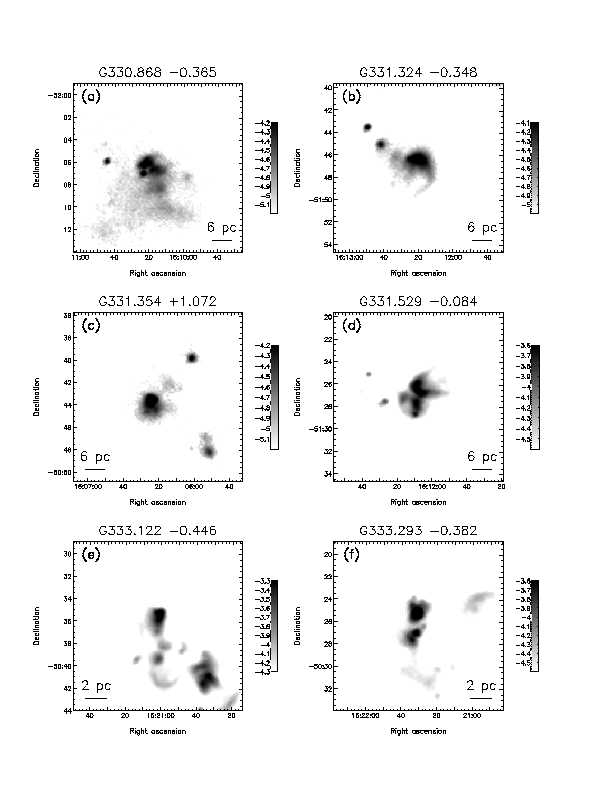}
\label{fig8}
\end{figure*}

\begin{figure*} {\bf Figure A9}. $MSX$ Band E images of GH\,{\sc ii} 
regions
\epsfysize=23cm   \epsfbox[0 -200 504 620]{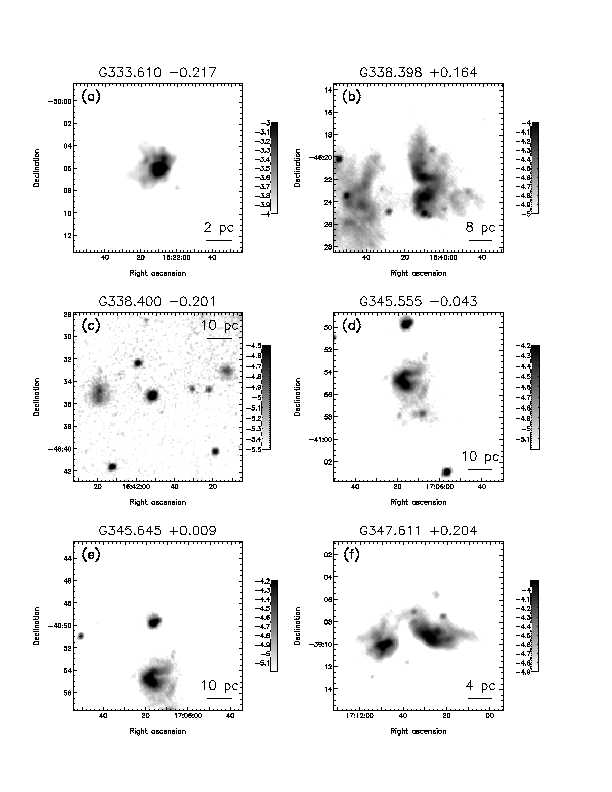}
\label{fig9}
\end{figure*}

\begin{figure*} {\bf Figure A10}. $MSX$ Band E images of Galactic and 
Local Group GH\,{\sc ii} regions.
\epsfysize=23cm   \epsfbox[0 -200 504 620]{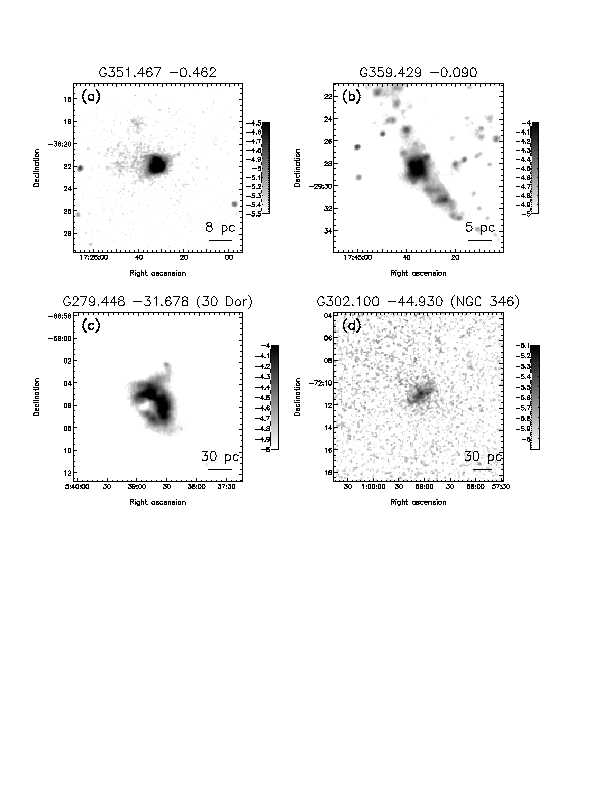}
\label{fig10}
\end{figure*}

\end{document}